\documentclass[10pt,twoside,twocolumn,english,prb, showpacs]{revtex4}
\usepackage[T1]{fontenc}
\usepackage[latin9]{inputenc}
\usepackage{amstext}
\usepackage{graphicx}
\usepackage{amssymb}
\usepackage{esint}

\makeatletter

\providecommand{\tabularnewline}{\\}

\@ifundefined{textcolor}{}
{%
 \definecolor{BLACK}{gray}{0}
 \definecolor{WHITE}{gray}{1}
 \definecolor{RED}{rgb}{1,0,0}
 \definecolor{GREEN}{rgb}{0,1,0}
 \definecolor{BLUE}{rgb}{0,0,1}
 \definecolor{CYAN}{cmyk}{1,0,0,0}
 \definecolor{MAGENTA}{cmyk}{0,1,0,0}
 \definecolor{YELLOW}{cmyk}{0,0,1,0}
 }

\usepackage{longtable}

\makeatother

\usepackage{babel}

\begin{document}

\title{Finite-temperature effects on the superfluid Bose-Einstein condensation
of confined ultracold atoms in three-dimensional optical lattices}

\author{T. P. Polak$^{1}$, T. K. Kope\'{c}$^{2}$}

\address{$^{1}$Adam Mickiewicz University of Pozna\'{n}, Faculty of Physics,
Umultowska 85, 61-614 Pozna\'{n}, Poland}

\address{$^{2}$Institute for Low Temperatures and Structure Research, Polish
Academy of Sciences, POB 1410, 50-950 Wroclaw 2, Poland}
\begin{abstract}
We discuss the finite-temperature phase diagram in the three-dimensional
Bose-Hubbard (BH) model in the strong correlation regime, relevant
for Bose-Einstein condensates in optical lattices, by employing a
quantum rotor approach. In systems with strong on site repulsive interactions,
the rotor U(1) phase variable dual to the local boson density emerges
as an important collective field. After establishing the connection
between the rotor construction and the the on--site interaction in
the BH model the robust effective action formalism is developed which
allows us to study the superfluid phase transition in various temperature--interaction
regimes. 
\end{abstract}


\maketitle

\section{Introduction}

The ability to confine ultracold quantum gases in optical lattices
is already having a major impact in fields as diverse as condensed-matter
physics and quantum information processing \cite{jaksch,greiner,fisher}.
An optical lattice is essentially a periodic intensity pattern that
is formed by the interference of two or more laser beams. The simplest
optical lattice consists of the region that is formed when two laser
beams with the same wavelength travelling in opposite directions meet
each other and form an interference pattern. This artificial structure
is able to trap an atom because the electric fields of the lasers
induce an electric dipole moment in the atom. However, the quantum-mechanical
tunnelling allows the atoms to spread through the optical lattice
to some degree. In a Bose-Einstein condensate, this tunnelling process
dominates the behaviour of the atoms, which causes the system to have
a perfect phase coherence between the matter waves on different lattices
sites. This quantum phase transitions of the Bose-Einstein condensates
loaded into the lowest vibrational level of single wells of an optical
lattice in the strict sense can exist only at temperature $T=0$.
However, in typical experimental situations we must take into consideration
thermal fluctuations in the particle number per site.In the presence
of the finite-temperature the nonzero value of the compressibility
is expected in contrast to the incompressible Mott state at $T=0$.
The experimental data only signals that the system nears a quantum
phase transition if the temperature is extrapolated to zero. What
the experiments really observe is a transition from the superfluid
to the normal liquid whose compressibility is very close to zero and
the system is practically a Mott insulator. This issue has gained
recently much attention \cite{buonsante,plimak,pupillo,gerbier}.
Theoretical \cite{kampf} and numerical \cite{prokofev,alet,caproso}
approaches have designed to these systems in three-dimensions but
only recently the finite-temperature effects have been studied systematically
\cite{yu,caproso,pollet}. Yu at. al. \cite{yu} presented study of
the finite-temperature behaviour of ultra-cold Bose atoms in three-dimensional
($3D$) optical lattices by the slave fermion and the slave boson
approaches to the Bose-Hubbard (BH) model. The finite-temperature
phase diagram was also investigated by Gerbier \cite{gerbier} in
the context of ultra-cold bosons confined in optical lattice in the
presence of an additional potential. Three regimes can be recognised
from the phase diagram: zero-temperature quantum phase, intermediate,
where Mott insulator (MI) features persist but superfluid (SF) region
is absent and the thermal region, where the Mott insulator properties
disappear. Despite of the few theoretical approaches to the problem
of the strongly interacting bosons at finite-temperatures many questions
still remain open and unsolved. Especially studies of the phase transitions
in the strongly correlated regime are scarce, where the repulsive
energy is the main energy scale in the system and we are far from
the limit of weakly interacting bosons. For these reasons, there is
still a strong need for approximate but robust treatments of strongly
correlated bosonic models in order to include the finite-temperature
properties especially relevant for phase diagrams for the $3D$ system
where temperature--induced phase transition exists. The purpose of
this paper is to present a robust theoretical description of correlated
bosonic systems which fulfils these goals. Our main idea is to focus
on the degrees of freedom associated to the relevant physical variable
associated to the Mott--superfluid transition, namely a quantum phase
field U(1) rotor field, which is dual to the local occupation number.
and acquires dynamic significance from the boson-boson interaction
\cite{kopec}.

\section{The model}

In experimental parameter regime the bosonic atoms with repulsive
interactions in a periodic lattice potential are perfectly described
by a Bose-Hubbard model which is the simplest nontrivial model describing
a bosonic many body system on a lattice which can not be mapped onto
a single particle problem. Nevertheless it captures essential effects
like a quantum phase transition from a superfluid state to a Mott
insulating state. The Hamiltonian of the model reads \begin{eqnarray}
\mathcal{H} & = & \frac{U}{2}\sum_{i}n_{i}^{2}-\sum_{\left\langle i,j\right\rangle }t_{ij}a_{i}^{\dagger}a_{j}-\bar{\mu}\sum_{i}n_{i},\label{hamiltonian2}\end{eqnarray}
 where, $a_{i}^{\dagger}$ and $a_{j}$ stand for the bosonic creation
and annihilation operators that obey the canonical commutation relations
$[a_{i},a_{j}^{\dagger}]=\delta_{ij}$, where $n_{i}=a_{i}^{\dagger}a_{i}$
is the boson number operator on the site $i$. Here, $\left\langle i,j\right\rangle $
identifies summation over the nearest-neighbour sites. Furthermore,
$t_{ij}$ is the hopping matrix element and describes the tunnelling
of bosons between neighbouring potential wells in the simple cubic
lattice and $\bar{\mu}/U=\mu/U+1/2$ is the shifted reduced chemical
potential which controls the number of bosons, $U>0$ is the on-site
repulsion. Due to the short range of the interactions compared to
the lattice spacing, the interaction energy is well described by this
term, which characterises a purely on-site interaction. The interaction
term tends to localise atoms to lattice sites. When the potential
depth of the optical lattice is increased, the tunnelling barrier
between neighbouring lattice sites is raised and the tunnelling matrix
element $t$ decreases. The on-site interaction $U$ on the other
hand is increased due to a tighter confinement of the wave function
for bosons on a lattice site. Therefore the ratio $U/t$ can be continuously
adjusted over a wide range by changing the strength of the lattice
potential. Finally we comment on the validity of the Hamiltonian in
Eq. (\ref{hamiltonian2}). The Bose-Hubbard model can be obtained
from many-body Hamiltonian with pseudo-potential interaction. However,
we must assume that thermal and mean interaction energy are much smaller
than the separation to the first excited band. Therefore, the exact
value of the temperature (in the limit where the recoil energy is
much smaller than maximum value of the lattice depth $E_{R}\ll V_{0})$
and can be calculated from the expression $k_{B}T\ll\hbar\omega_{0}$
where $\hbar\omega_{0}$ is the energy separated a number of vibrational
levels in an infinite periodic potential. The energy to comparison
can be taken from single site of the lattice. Moreover the hopping
matrix elements are nonzero or non-negligible only to nearest-neighbours,
so there are always negative in the lowest band.

\section{Phase action and order parameter}

\begin{figure}
\includegraphics[scale=0.48]{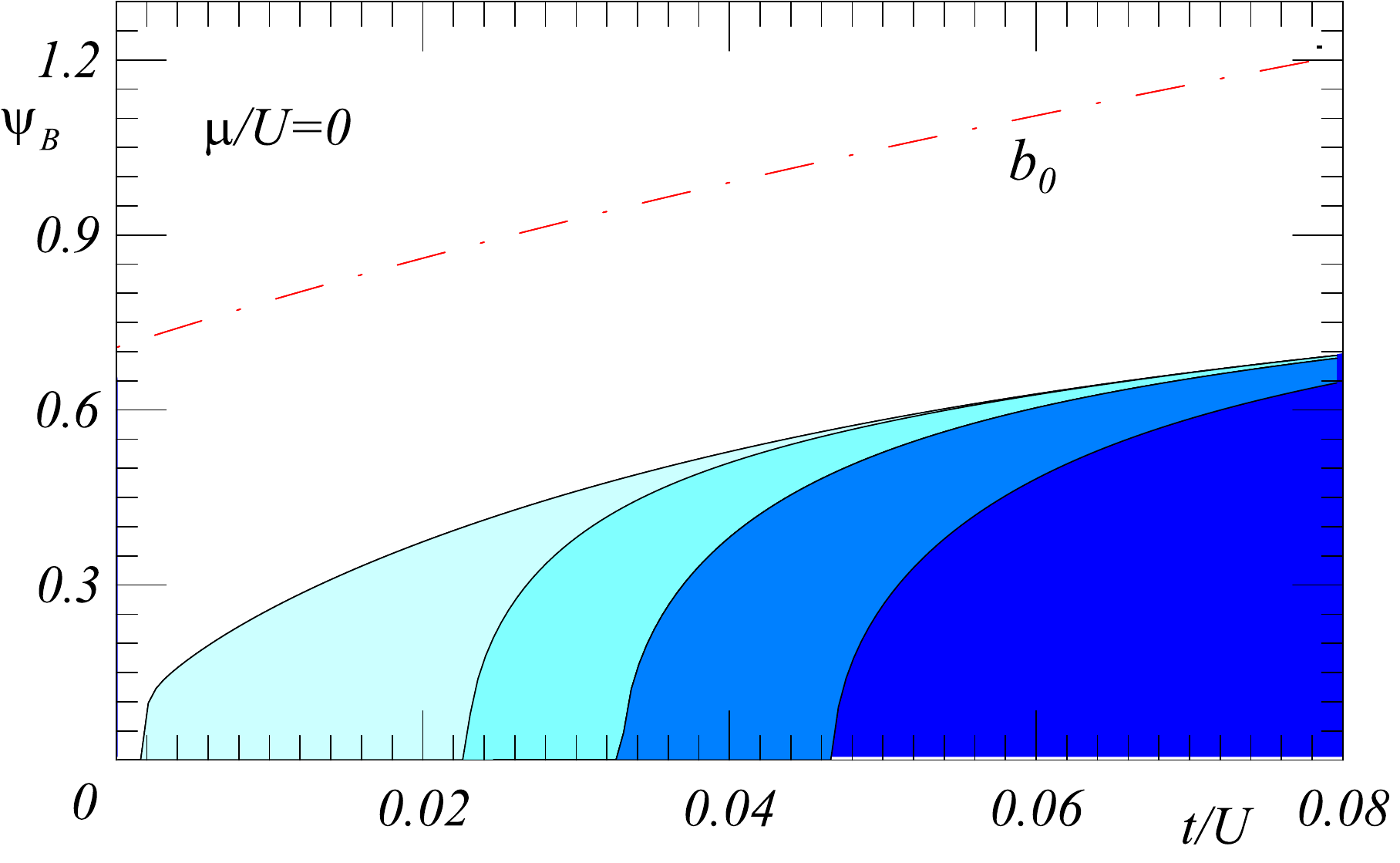} \caption{The superfluid order parameter $\psi_{B}$ measuring the degree of
the phase coherence as a function of the interaction strength $t/U$
for several values of the temperature $k_{B}T/U=0.00,0.05,0.10,0.20$
from the left to the right and fixed chemical potential. The dash--dotted
line is the corresponding amplitude amplitude $b_{0}$ (see, Eq.\ref{ppo}
). \label{superfluid-order}}

\end{figure}

The functional integral representation of models for correlated bosons
allows us to implement efficiently the method of treatment. The partition
function is written in the form\begin{equation}
\mathcal{Z}=\int\left[\mathcal{D}\bar{a}\mathcal{D}a\right]e^{-\mathcal{S}\left[\bar{a},a\right]}\end{equation}
 and the bosonic path integral is taken over the complex fields $a_{i}\left(\tau\right)$
with the action $\mathcal{S}$ given by \begin{equation}
\mathcal{S}[\bar{a},a]=\sum_{i}\int_{0}^{\beta}d\tau\left[\bar{a}_{i}\left(\tau\right)\frac{\partial}{\partial\tau}a_{i}\left(\tau\right)+\mathcal{H\left(\tau\right)}\right],\label{act}\end{equation}
 where $\beta=1/k_{\mathrm{B}}T$ and $T$ is the temperature. Since
Hamiltonian is not quadratic in the fields $a_{i}$ we have to decouple
first the interaction term in Eq. (\ref{hamiltonian2}) by means of
a Gaussian integration over the auxiliary scalar potential fields
$V_{i}\left(\tau\right)$ which periodic part $V_{i}^{P}\left(\tau\right)$
couples to the local particle number through the Josephson-like relation
$\dot{\phi}_{i}\left(\tau\right)=V_{i}^{P}\left(\tau\right)$ where
$\dot{\phi}_{i}\left(\tau\right)\equiv\partial\phi_{i}\left(\tau\right)/\partial\tau$.
The phase field satisfies the periodicity condition $\phi_{i}\left(\beta\right)=\phi_{i}\left(0\right)$
as a consequence of the periodic properties of the $V_{i}^{P}\left(\tau\right)$
field. Next, we perform the local gauge transformation to the new
bosonic variables \begin{equation}
a_{i}\left(\tau\right)=b_{i}\left(\tau\right)\exp\left[i\phi_{i}\left(\tau\right)\right].\end{equation}
 Using such a description is justified by the definition of the order
parameter \begin{equation}
\Psi_{B}\equiv\left\langle a_{i}\left(\tau\right)\right\rangle =\left\langle b_{i}\left(\tau\right)\exp\left[i\phi_{i}\left(\tau\right)\right]\right\rangle =b_{0}\psi_{B}\label{ppo}\end{equation}
 which non-vanishing value signals a macroscopic quantum phase coherence
(in our case we identify it as SF state). The system can be then described
by a macroscopic wave function since the many-body state is a product
over identical single particle states. Therefore a macroscopic phase
is well defined on each lattice site and the system is superfluid.
On the other hand the atom number per site is uncertain , and therefore
one would find a random atom number in a measurement. In the large
$U$ limit the amplitude $b_{0}\equiv\langle b_{i}\rangle$ (see Fig.
\ref{superfluid-order}) has a nonzero value, but to achieve the superfluidity,
the phase variables must also become stiff and, in consequence, $\psi_{B}\equiv\exp\left[i\phi_{i}(\tau)\right]\neq0$.
Furthermore, we parametrise the boson fields $b_{i}\left(\tau\right)=b_{0}+b_{i}^{'}\left(\tau\right)$
and restrict our calculations to the phase fluctuations dropping the
amplitude dependence\cite{polak} which is justified in the large
$U/t$ limit. The $\mathrm{U}\left(1\right)$ group governing the
phase field is compact and $\phi\left(\tau\right)$ has the topology
of a circle, so that instanton effects can arise due to non-homotopic
mappings of the configuration space onto the gauge group $\mathrm{U}\left(1\right)$.
Accordingly, the path integral reads \begin{equation}
\int\left[\mathcal{D}\phi\right]...\equiv\sum_{\left\{ m_{i}\right\} }\prod_{i}\int_{0}^{2\pi}d\phi_{i}\left(0\right)\int_{_{\phi_{i}\left(0\right)}}^{\phi\left(\tau\right)_{i}+2\pi m_{i}}[{\mathcal{D}}\phi_{i}\left(\tau\right)]...\end{equation}
 which is performed by taking phase configurations that satisfy boundary
condition $\phi_{i}\left(\beta\right)-\phi_{i}\left(0\right)=2\pi m_{i}$
($m_{i}=0,\pm1,\pm2,\dots$), where the winding numbers $n_{i}$ label
the distinct homotopy classes of the $\mathrm{U}\left(1\right)$ group.
Thus the paths can be divided into topologically distinct classes,
characterised by a winding number defined as the net number of times
the world line wraps around the system in the {}``imaginary time''
direction. Integrating the action in Eq. (\ref{act}) over the bosonic
fields we obtain the effective Lagrangian in terms of the phase-only
variables \begin{eqnarray}
\mathcal{S}_{\mathrm{ph}}\left[\phi\right] & = & \int_{0}^{\beta}d\tau\left\{ \sum_{i}\left[\frac{1}{2U}\dot{\phi_{i}^{2}}\left(\tau\right)+\frac{1}{i}\frac{\bar{\mu}}{U}\dot{\phi_{i}}\left(\tau\right)\right]\right.\nonumber \\
 &  & \left.-\sum_{\left\langle i,j\right\rangle }e^{\phi_{i}\left(\tau\right)}J_{ij}e^{-\phi_{j}\left(\tau\right)}\right\} ,\label{action only phase}\end{eqnarray}
 with the phase stiffnesses $J_{ij}=b_{0}^{2}t_{ij}$, where the amplitude
$b_{0}^{2}=\left(\sum_{\left\langle i,j\right\rangle }t_{ij}+\bar{\mu}\right)/U$
originates from the saddle point condition \begin{equation}
\left.\frac{\partial\mathcal{S}[\bar{b},b]}{\partial b}\right|_{b=b_{0}}=0.\end{equation}
The result of the gauge transformations is that we have managed to
cast the strongly correlated problem into a system of mutually noninteracting
effective bosons, submerged in the bath of strongly fluctuating $\mathrm{U}\left(1\right)$
phase fields, whose dynamics is governed by the energy scale set by
the on-site interaction $U$ that drives the Mott transition.

\section{Treatment of the action of quantum rotors}

Now, we devise a systematic way of treatment for the fluctuating phase
fields contained in the action in Eq. (\ref{action only phase}) that
enables us to obtain an effective non-linear sigma-field theory that
respects the symmetry properties of the model and satisfies the Mermin-Wagner
theorem, thereby improving the pure mean-filed approach known from
its restricted ability to deal with the spatial fluctuations. To proceed,
it is convenient to replace the phase degrees of freedom by the complex
field $\psi_{i}\equiv e^{\phi_{i}\left(\tau\right)}$ which satisfies
the periodic boundary condition $\psi_{i}\left(\beta\right)=\psi_{i}\left(0\right)$.
This can be done by implemented the Fadeev-Popov method with the Dirac
delta functional resolution of the unity:\cite{kopec1}\begin{eqnarray}
1 & \equiv & \int\left[\mathcal{D}\bar{\psi}\mathcal{D}\psi\right]\delta\left(\sum_{i}\left|\psi_{i}\left(\tau\right)\right|^{2}-N\right)\nonumber \\
 & \times & \prod_{i}\delta\left(\psi_{i}-e^{i\phi_{i}\left(\tau\right)}\right)\delta\left(\bar{\psi}_{i}-e^{-i\phi_{i}\left(\tau\right)}\right),\label{popov}\end{eqnarray}
 where we take $\psi_{i}$ as continuous variable but constrained
(on the average) to have the unimodular value. We can solve the constraint
by introducing the Lagrange multiplier $\lambda$ which adds the quadratic
terms (in the $\psi_{i}$ fields) to the action Eq. (\ref{action only phase}).
The partition function is written in form \begin{eqnarray}
\mathcal{Z} & = & \int_{-i\infty}^{+i\infty}\left[\frac{{\mathcal{D}}\lambda}{2\pi i}\right]e^{-N\beta\mathcal{F}\left(\lambda\right)},\end{eqnarray}
 where the free energy per site $\mathcal{F}=-\ln\mathcal{Z}/\beta N$
is given by: \begin{eqnarray}
\mathcal{F} & = & -\lambda-\frac{1}{N\beta}\ln\int\left[\mathcal{D}\bar{\psi}\mathcal{D}\psi\right]e^{-{\mathcal{S}}_{\mathrm{eff}}[\bar{\psi},\psi]}\nonumber \\
{\mathcal{S}}_{\mathrm{eff}}[\bar{\psi},\psi] & = & \sum_{\langle i,j\rangle}\int_{0}^{\beta}d\tau d\tau^{'}\left[\left(J_{ij}+\lambda\delta_{ij}\right)\delta\left(\tau-\tau'\right)\right]\nonumber \\
 &  & \left.-\mathcal{\gamma}_{ij}\left(\tau,\tau'\right)\right]\bar{\psi}_{i}(\tau)\psi_{j}(\tau'),\label{free energy}\end{eqnarray}
 and $\gamma_{ij}\left(\tau,\tau'\right)=\left\langle \exp\left\{ -i\left[\phi_{i}\left(\tau\right)-\phi_{j}\left(\tau'\right)\right]\right\} \right\rangle $
is the two-point phase correlator associated with the order parameter
field, where $\left\langle \dots\right\rangle $ is the averaging
with respect to the action in Eq. (\ref{action only phase}). 
\begin{figure}
\includegraphics[scale=0.4]{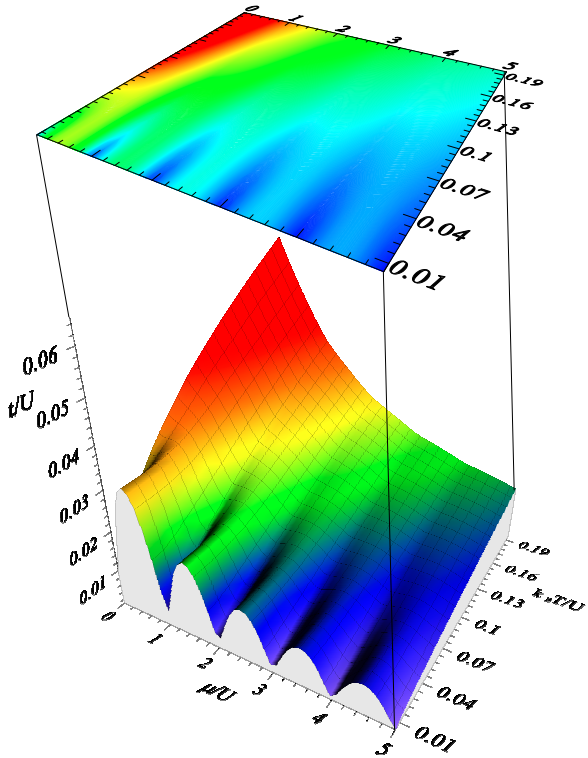} \caption{Phase diagram for BH model on a cubic lattice as a function of chemical
potential and temperature. The Mott insulator phase is found only
at $T=0$ within each lobe of integer boson density since for T=0
this state is incompressible.; at $T>0$ a region with finite compressibility
emerges (due to the thermal activation), so we have here the {}``disordered
state\textquotedbl{} without phase coherence. Above the critical surface
the superfluid region takes place. Top panel: the projection of the
critical surface onto the plane temperature--chemical potential in
a form of the density plot. \label{Finite-temperature-phase-diagram}}

\end{figure}

The action with the the topological contribution, after Fourier transform,
we write as \begin{equation}
\mathcal{S}_{\mathrm{eff}}[\bar{\psi},\psi]=\frac{1}{N\beta}\sum_{\mathbf{k},\ell}\bar{\psi}_{\mathbf{k},\ell}\mathrm{\Gamma}_{\mathbf{k}}^{-1}\left(\omega_{\ell}\right)\psi_{\mathbf{k},\ell},\end{equation}
 where $\mathrm{\Gamma}_{\mathbf{k}}^{-1}\left(\omega_{\ell}\right)=\lambda-J_{\mathbf{k}}+\gamma^{-1}\left(\omega_{\ell}\right)$
is the inverse of the propagator and $\omega_{\ell}=2\pi\ell/\beta$
($\ell=0,\pm1,\pm2,\dots)$ stands for the Bose-Matsubara frequency,
while the phase correlator, after Fourier transform, can be written
as: \begin{equation}
\gamma\left(\omega_{\ell}\right)=\frac{1}{\mathcal{Z}_{0}}\frac{4}{U}\sum_{m=-\infty}^{+\infty}\frac{e^{-\frac{1}{2}\beta U\left(m+\frac{\bar{\mu}}{U}\right)^{2}}}{1-4\left(m+\frac{\bar{\mu}}{U}-\frac{i\omega_{\ell}}{U}\right)^{2}},\label{correlator}\end{equation}
 where $\mathcal{Z}_{0}=\sum_{m=-\infty}^{+\infty}\exp\left[-\frac{1}{2}\beta U\left(m+\frac{\bar{\mu}}{U}\right)^{2}\right]$
is the partition function for the set of non-interacting quantum rotors.
Within the phase coherent state the order parameter is given by \begin{equation}
1-\psi_{B}^{2}=\frac{1}{N\beta}\sum_{\mathbf{k},\ell}\frac{1}{\lambda_{0}-J_{\mathbf{k}}+\gamma^{-1}\left(\omega_{\ell}\right)}.\label{critical line}\end{equation}
 For the simple cubic lattice we write $J_{\mathbf{k}}=\left(12t+\bar{\mu}\right)t_{\mathbf{k}}/U$
with the dispersion $t_{\mathbf{k}}=2t\left(\cos k_{1}+\cos k_{2}+\cos k_{3}\right)$.
The phase boundary is determined by the divergence of the order parameter
susceptibility $\Gamma_{\mathbf{k}=0}\left(\omega_{\ell=0}\right)=\infty$,
which determines the critical value of the Lagrange parameter $\lambda=\lambda_{0}$
that stays constant in the whole ordered phase. After summation over
Matsubara frequency the superfluid state order parameter becomes \begin{eqnarray}
1-\psi_{B}^{2} & = & \frac{1}{4N}\sum_{\mathbf{k}}\frac{1}{{\Lambda_{\textbf{k}}}}\left\{ \coth\left[\frac{1}{2}\beta U\left({\Lambda_{\textbf{k}}}-\upsilon\left(\frac{\mu}{U}\right)\right)\right]\right.\nonumber \\
 & + & \left.\coth\left[\frac{1}{2}\beta U\left({\Lambda_{\textbf{k}}}+\upsilon\left(\frac{\mu}{U}\right)\right)\right]\right\} .\label{critical line final}\end{eqnarray}
 In the above equation $\Lambda_{\textbf{k}}^{2}=\left(J_{0}-J_{\mathbf{k}}\right)/U+\upsilon^{2}\left(\mu/U\right)$
and $\upsilon\left(\mu/U\right)=\mathrm{frac}\left(\mu/U\right)-1/2,$
where $\mathrm{frac}\left(x\right)=x-\left[x\right]$ is the fractional
part of the number and $\left[x\right]$ is the floor function which
gives the greatest integer less then or equal to $x$.

\section{Results}

When the strength of the interaction term relative to the tunnelling
term in the Bose-Hubbard Hamiltonian is changed, the system reaches
a quantum critical point in the ratio of $U/t$, for which the system
will undergo a quantum phase transition from the superfluid ground
state to the Mott insulator ground state. In three dimensions, according
to Mermin-Wagner theorem this phase transition can occur also at non-zero
temperature. The finite-temperature phase diagram of the model can
be calculated from Eq. (\ref{critical line final}) by introducing
the density of states for simple cubic lattice in order to perform
the sum over the lattice wave vectors: \begin{eqnarray}
\rho\left(\xi\right) & = & \frac{1}{\pi^{3}t}\int_{a_{1}}^{a_{2}}\frac{d\epsilon}{\sqrt{1-\epsilon^{2}}}\Theta\left(1-\frac{\left|\xi\right|}{3t}\right)\nonumber \\
 & \times & \mathbf{K}\left(\sqrt{1-\left(\frac{\xi}{2t}+\frac{\epsilon}{2}\right)^{2}}\right)\end{eqnarray}
 with $a_{1}=\mathrm{min}\left(-1,-2-\xi/t\right)$ and $a_{2}=\mathrm{max}\left(1,2-\xi/t\right)$;
$\mathbf{K}\left(x\right)$ is the elliptic function of the first
kind \cite{abramovitz}.

\begin{figure}
\includegraphics[scale=0.45]{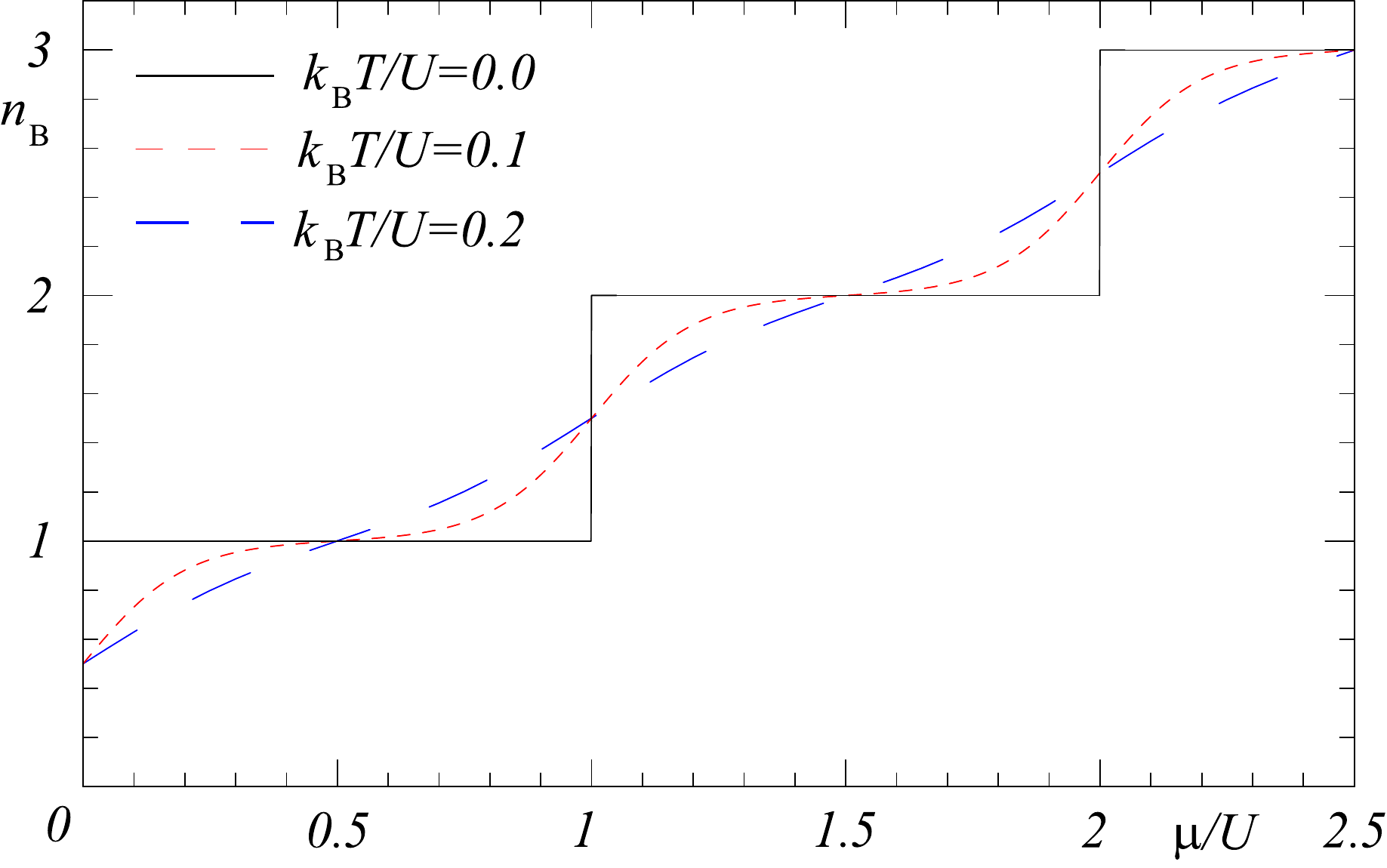} \caption{Bosonic occupation number $n_{B}$ as a function of the chemical potential,
for several values of the temperature as indicated in the plot calculated
for the extreme interaction limit ($U/t=\infty$, to highlight the
sole temperature effect). \label{Bosonic-occupation-number}}

\end{figure}

\begin{figure}
\includegraphics[scale=0.65]{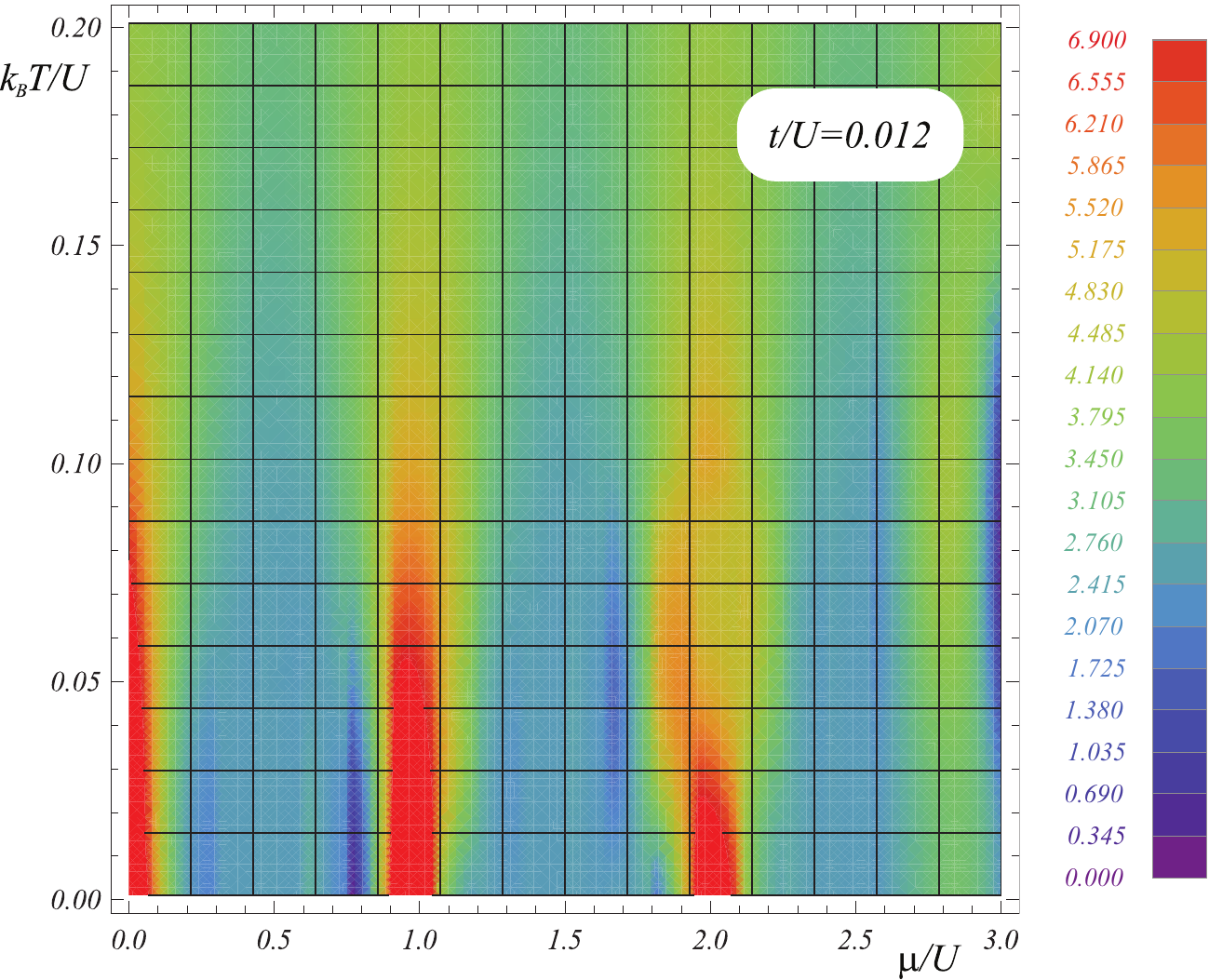} \caption{Density plot of the normalised compressibility $\tilde{\kappa}=\kappa/U$,
where $\kappa=\partial n_{B}/\partial\mu$ calculated numerically
as a function of temperature and chemical potential for $t/U=0.012$.
Dark (red) shading around the integer values of $\mu/U$ corresponds
to high values of $\kappa$ whereas the light (blue--to--green) shading
marks the region of diminishing compressibility.\label{compressibility}}

\end{figure}

\begin{figure}
\includegraphics[scale=0.8]{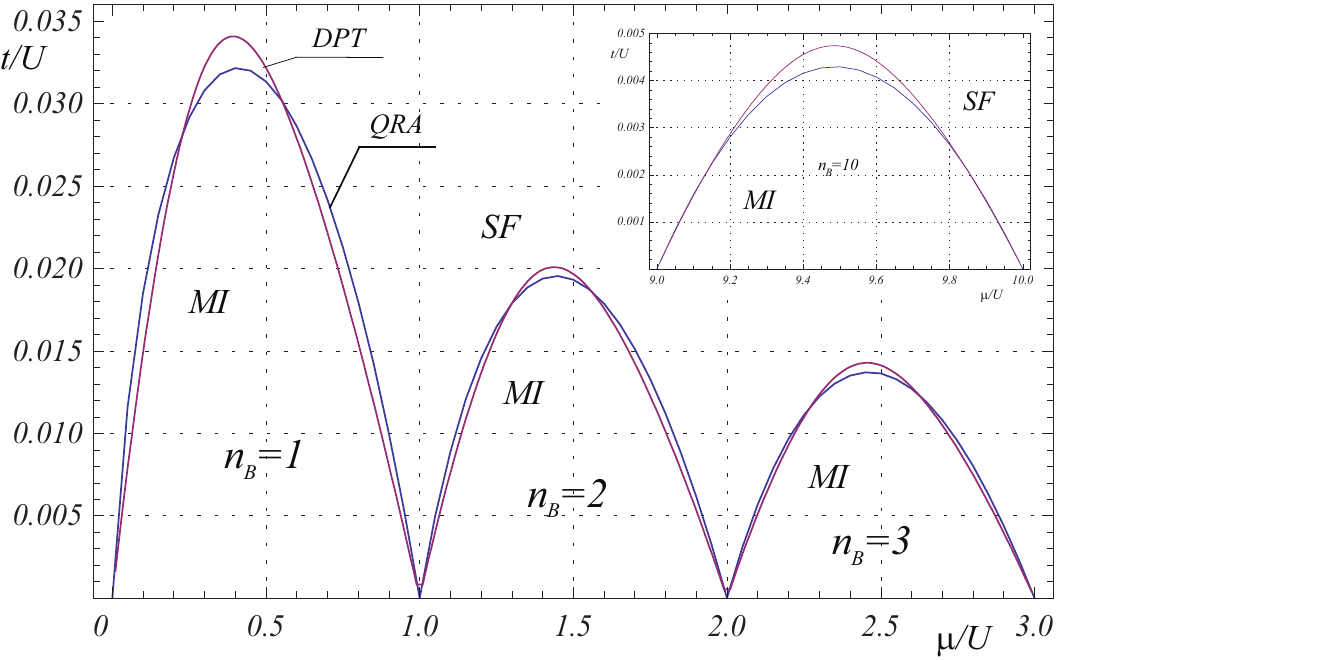} \caption{The comparison of the zero-temperature phase diagram calculated from
the diagrammatic perturbation theory\cite{teichmann} (DPT) and our
results obtained in the frame of quantum rotor approach (QRA) with
$n_{B}=1$, $2$, $3$ and $10$. We found the critical value $\left(t/U\right)_{\mathrm{crit}}$
(see also Table \ref{comparison}) for the tip of the $n$th Mott-insulator
lobe is always slightly lower than obtained from DPT and Monte-Carlo
simulations \cite{capogrosso-sansone}. \label{Bosonic-comp}}

\end{figure}

A lobe-like structure (see Fig. \ref{Finite-temperature-phase-diagram}),
similar to the zero-temperature case, becomes flat with increasing
temperature. The lobes with a larger boson occupation number are more
stable against temperature. The stability comes from higher values
of the repulsive energy $U$. Therefore, at temperature $T=0$ the
interaction in the system \cite{polak} governs the quantum phase
transition. By decreasing the value of the repulsive energy we can
achieve superfluid phase. In real physical realizations of the BH
model thermal excitations are always present and also can activate
a phase transition. We see it clearer by calculating a bosonic occupation
number. Decomposing the phase field in terms of a periodic field and
linear in $\tau$ term we calculate the effects of the fixed boson
number $n_{B}=N^{-1}\sum_{i}\left\langle \bar{a}_{i}\left(\tau\right)a_{i}\left(\tau\right)\right\rangle $
in the system. The total boson density $n_{B}=n_{b}+\delta n_{b}$
consists of the occupation number for neutral bosons $n_{b}$ and
a contribution $\delta n_{b}$ from a fluctuating phase field. For
$T=0$ we recognise a steps of fixed integer filling of bosons (see
Fig. \ref{Bosonic-occupation-number}). With increasing temperature,
typical for the Mott state, steps-like profile becomes smoother. Therefore,
bosons placed in the Mott state get energy required to move from one
lattice site to another from thermal fluctuations. The temperature
$k_{\mathrm{B}}T/U\sim0.2$, where the occupation number characteristic
becomes flat, is similar to recognised as a melting temperature for
the condensate slowly loaded into the optical potential in the presence
of an smoothly varying trap \cite{gerbier}. The increasing value
of the thermal energy, in analogy to the decreasing repulsive interaction
\cite{polak}, can lead to the situation where the sharp steps of
the MI state become indistinct (see Fig. \ref{Bosonic-occupation-number}).
The Mott-insulator to superfluid quantum phase transition is rigorously
present only at zero temperature, whereas at finite-temperature thermal
fluctuations induce a phase transition between a superfluid and a
normal phase. However, at sufficiently low temperatures, a remnant
of the insulating phase still persists within the normal phase. In
these conditions it is possible to observe a sharp crossover between
a compressible normal fluid and a phase characterised by a vanishing
compressibility (see, Fig. \ref{compressibility}). Regarding the
comparison of our method with the previous approaches we found that
our results are in good agreement with other calculations including
numerical quantum Monte-Carlo \cite{capogrosso-sansone}, diagrammatic
perturbation theory\cite{teichmann}) and analytical works based on
the strong coupling perturbation theory \cite{freericks}, see Table
\ref{comparison} and Fig. \ref{Bosonic-comp}.

%
\begin{table}
\caption{\label{comparison} Comparison of the maximum of the critical value
for $t/U$ parameter (as a function of the normalised chemical potential
$\mu/U$) at the tip of the first ($n_{B}=1$) MI lobe for square
lattice with several numerical (QMC - quantum Monte-Carlo \cite{capogrosso-sansone},
DPT - diagrammatic perturbation theory \cite{teichmann}) and analytical
work: SCPT - strong coupling perturbation theory \cite{freericks}.
QRA - our calculations using quantum rotor approach).}
\begin{tabular}{c|c|c|c|c}
 & QMC  & DPT  & PA  & QRA\tabularnewline
\hline 
$t/U$  & $0.03408(2)$  & $0.03407$  & $0.034737$  & $0.03215$\tabularnewline
$\mu/U$  & $0.389$  & $0.393$  & $0.37905$  & $0.41$\tabularnewline
\end{tabular}
\end{table}

\section{Final remarks and outlook}

To conclude, in this paper we have presented a study of the finite-temperature
transition of the three-dimensional Bose-Hubbard model relevant for
Bose-Einstein condensates trapped in three-dimensional optical lattice
potentials which allow to enter a new regime in the many body physics
of ultracold atomic gases. Ultracold atoms in optical lattice potentials
represent a rather perfect realization of the Bose-Hubbard model with
a high degree of control. Parameters like the ratio between the on-site
interaction and tunnel coupling or the filling factor can be widely
adjusted, and observables like long range phase coherence or number
statistics can be measured. In order to approach these issues theoretically
we employed the $\mathrm{U}\left(1\right)$ quantum rotor approach
and a path integral formulation of quantum mechanics including a summation
over a topological charge, explicitly tailored for the BH Hamiltonian.
This method can give the thermodynamics of the Bose-Hubbard model
in the limit of strong interactions. Our aim was then to analyse the
phase transitions that may occur in such system at finite-temperature,
and determine the general features of the associated phase diagrams.
We demonstrated the evolution of zero temperature Mott lobes and Mott
plateaus when the temperature is increasing. The technique used in
this paper can be conveniently extended to more general situations,
including i.e. multi-species bosonic systems. Hamiltonians other than
the pure Bose-Hubbard Hamiltonian could be realized and studied by
the method outlined in three-dimensional optical lattice potentials.
For example by using a multi component gas and interspecies Feshbach
resonances, an intriguing system could be created. Furthermore, it
should be possible to consider other interactions as effective next
neighbour interaction, dipolar interactions or spin interaction due
to spin dependent tunnelling. Other generalisations of the on site
Hubbard model are of course possible by including the influence of
the disorder in the description.

We would like to thank N. Teichmann for an access to the numerical
data needed for comparison. One of us (T.K.K) acknowledges the support
by the Ministry of Education and Science MEN under Grant No. 1 P03B
103 30 in the years 2006-2008.

\end{document}